\documentclass[11pt]{article}
\usepackage[margin=0.75in]{geometry}
\usepackage{graphicx}
\usepackage{float}
\usepackage{amsfonts}
\usepackage{amsmath,amsthm,amssymb}
\usepackage{amsmath,amssymb}
\usepackage{epstopdf}
\usepackage{subfigure}
\usepackage{algorithm, algorithmic}
\usepackage[font=footnotesize]{caption}
\usepackage{color}
\usepackage{wrapfig}
\usepackage{marvosym}
\usepackage{cleveref}
\usepackage{ragged2e}
\usepackage[scaled]{helvet} 
\usepackage{fourier}
\usepackage{pdflscape}
\usepackage[english]{babel}
\usepackage{multirow}
\usepackage[usenames,dvipsnames,svgnames,table]{xcolor}

\newcommand{\R}{\mathbb{R}}

\newcommand{\bx}{\boldsymbol{x}}

\newcommand{\by}{\boldsymbol{y}}
\newcommand{\bR}{\boldsymbol{R}}

\newcommand{\bn}{{\boldsymbol{n}}}
\newcommand{\ba}{\boldsymbol{a}}
\newcommand{\bb}{\boldsymbol{b}}

\newcommand{\bv}{\boldsymbol{v}}
\newcommand{\bu}{\boldsymbol{u}}

\newcommand{\bLambda}{\boldsymbol{\Lambda}}

\newcommand{\bPhi}{\boldsymbol{\Phi}}

\newcommand{\supp}{\mathrm{supp}}

\DeclareMathOperator{\sign}{sign}
\DeclareMathOperator*{\argmin}{arg\,min}

\begin{document}

\title{Methods for Quantized Compressed Sensing}

\author{Hao-Jun Michael Shi\thanks{University of California, Los Angeles}, Mindy Case$^*$, Xiaoyi Gu$^*$, Shenyinying Tu$^*$, Deanna Needell\thanks{Claremont McKenna College}
}

\maketitle

\begin{abstract}
In this paper, we compare and catalog the performance of various greedy quantized compressed sensing algorithms that reconstruct sparse signals from quantized compressed measurements. We also introduce two new greedy approaches for reconstruction: Quantized Compressed Sampling Matching Pursuit (QCoSaMP) and Adaptive Outlier Pursuit for Quantized Iterative Hard Thresholding (AOP-QIHT). We compare the performance of greedy quantized compressed sensing algorithms for a given bit-depth, sparsity, and noise level.
\end{abstract}



\section{Introduction} \label{intro}
Compressed sensing (CS) is an emergent linear sampling framework that enables reconstruction of sparse signals from a small number of linear measurements relative to the total dimension of the signal space. In particular, given the acquired compressed signal $\by \in \mathbb{R}^M$ and measurement matrix $\bPhi \in \mathbb{R}^{M \times N}$, one seeks to reconstruct the signal $\bx \in \mathbb{R}^N$ by solving the (possibly noisy) underdetermined linear system
$
\by = \bPhi \bx.
$
Cand\`{e}s, et. al. \cite{candes2006stable} demonstrated that $K$-sparse signals, i.e. $\bx$ that satisfy $\| \bx \|_0 = |\text{supp}(\bx)| \leq K$, may be robustly reconstructed by an $\ell_1$-minimization program if $\bPhi$ satisfies the restricted isometry property (RIP) \cite{RefWorks:48}. Random matrices whose entries are chosen according to an appropriately chosen i.i.d. distribution and random submatrices of structured matrices have been shown to satisfy the RIP with high probability \cite{RefWorks:285}.

However, classic compressive sensing assumes that the measurements are continuous and real-valued. In practice, real measurements must be \textit{quantized}, or mapped to a discrete value from some finite set. In addition, in real-world applications, severe quantization may be preferred since low-bit measurements tend to be more efficient and inexpensive in acquisition and robust to amplification and other errors. Since compressed sensing seeks to store signal information in a compressed state and memory is often measured in bits, this motivates the rigorous treatment of quantization in compressed sensing. 

We define quantization as follows: given some $x \in \mathbb{R}$, our \textit{quantizer} or \textit{quantization function} is defined as
\begin{equation*}
f_Q(x) = \begin{cases}
m_1 & \mbox{ if } x \in (-\infty, \tau_2)\\
m_i & \mbox{ if } x \in [\tau_i, \tau_{i + 1} ) \text{ for } i = 2, ..., Q,
\end{cases}
\end{equation*}
where $\{\tau_1 = -\infty, \tau_2, ..., \tau_{Q + 1} = \infty \}$ is a partition of the real line and $\{m_1, ..., m_Q\}$ consist of quantization values. When $f_Q$ is applied to a vector $\bx$, we simply quantize each component of $\bx$ appropriately. This admits the quantized compressed sensing framework:
\begin{equation*}
\by = f_Q(\bPhi \bx),
\end{equation*}
where we would like to solve for the signal $\bx \in \mathbb{R}^n$. An extreme case of quantization is the 1-bit quantizer, where
\begin{equation*}
f_Q(\bPhi \bx) = \text{sign}(\bPhi \bx).
\end{equation*}
Since the magnitude of the signal is lost through low-bit or 1-bit quantization, we may restrict our sparse signals to the hypersphere
\begin{equation*}
S^{N - 1} = \{ \bx \in \mathbb{R}^N : \| \bx \|_2 = 1\}.
\end{equation*}

 We thus seek to solve the optimization problem:

\begin{equation}\label{eq:l0}
\begin{aligned}
\min_{\bx\in S^{N-1}} \| \bx\|_0 \quad \text{s.t.} \quad f_Q(\bPhi \bx) = \by.
\end{aligned}
\end{equation}

Equivalently, we can define the \textit{quantization region}
\begin{equation*}
\mathcal{R}_{\by} := \mathcal{R}_{y_1} \times ... \times \mathcal{R}_{y_M}
\end{equation*}
where $\mathcal{R}_{y_i}$ is the quantization region for the $i$th component of $\by$, or the interval $f_Q^{-1}(y_i)$. 

This gives the equivalent formulation to the above which we consider:

\begin{equation}\label{eq:l0a}
\begin{aligned}
\min_{\bx\in S^{N-1}} \| \bx\|_0 \quad \text{s.t.}
\quad \bPhi \bx \in \mathcal{R}_{\by}.
\end{aligned}
\end{equation}

The 1-bit variant of this compressed sensing problem was initially introduced and studied by P. Boufounos, et. al. \cite{RefWorks:6}, which then led to the development of many subsequent algorithms for 1-bit reconstruction (see e.g. \cite{biht,plan2013one,baraniuk2014exponential,RefWorks:19}).  One of the first was Binary Iterative Hard Thresholding (BIHT) which demonstrated accurate recovery \cite{biht}. Some variants of BIHT were then introduced to account for noise from acquisition and transmission, including Adaptive Outlier Pursuit (AOP) \cite{RefWorks:19}.
In a more general setting, methods for quantized compressed sensing have also been introduced (see e.g. \cite{dai2009distortion,dai2011information,jacques2013quantized}).

{\bfseries Contribution.}  We study several greedy approaches to solving this problem, and catalog precisely the tradeoff between reconstruction error, bit depth, and number of measurements.  We believe such a catalog is useful for guiding practitioners in selecting which method to use in quantized CS, as in \cite{blanchard2015performance} for the classical case.  In addition, we develop new approaches which outperform existing methods in some contexts.

{\bfseries Organization.} We introduce existing methods for 1-bit and quantized CS in Sections \ref{basis pursuit} and \ref{current greedy}.  Then, in Section \ref{novel greedy} we motivate and propose two new adaptations which we show outperform existing methods in certain regimes in Section \ref{numerical experiments}.  Also in this section we perform extensive experiments which catalog the reconstruction behavior for our proposed methods and existing methods.  We do so in such a way that given a bit-budget, signal size, and sparsity level, one can optimally select the algorithm for best reconstruction error.  Lastly, we conclude in Section \ref{conclusion}.

\section{Quantized Basis Pursuit} \label{basis pursuit}

In classical compressed sensing, a relaxation from the $\ell_0$-norm to the $\ell_1$-norm is used to obtain the following basis pursuit formulation:
\begin{equation*}
\begin{aligned}
\min_{\bx} \| \bx\|_1 \quad \text{s.t.} \quad \bPhi \bx = \by,
\end{aligned}
\end{equation*}
which has provable robust reconstruction guarantees
\cite{RefWorks:48,candes2006stable}. 

Similarly, W. Dai, et. al. \cite{dai2011information} proposed a quantized basis pursuit approach. Given a quantizer $f_Q$, one defines vectors $\bb_1$ and $\bb_2$ consisting of the thresholds for each coordinate's quantization region. That is, we may solve
\begin{equation*}
\begin{aligned}
\min_{\bx}\| \bx\|_1 \quad \text{s.t.} \quad \bb_1 \leq \bPhi \bx \leq \bb_2.
\end{aligned}
\end{equation*}

Since both of these problems are linear programs, one can use traditional interior-point, simplex methods, or even Bregman algorithms to solve them.  However, these algorithms are typically slower, motivating the development of greedy approaches.

\section{Current Greedy Methods}\label{current greedy}

We first present several current greedy algorithms for quantized compressed sensing found in the literature. 

\subsection{Quantized Subspace Pursuit (QSP)}

The Subspace Pursuit (SP) algorithm was developed by W. Dai, et. al. \cite{dai2009subspace} for classic compressed sensing reconstruction. In particular, SP has provable reconstruction capability similar to basis pursuit methods. It was adapted for quantized compressed sensing in \cite{dai2011information} as we describe here.

Let $T \subset \{1, ..., N\}$ be an index set and denote $\bPhi_T$ and $\bx_T$ as the truncated matrix consisting of the columns of $\bPhi$ indexed by $T$ and entries of $\bx$ indexed by $T$, respectively. 

We can define the set
\begin{equation*}
\mathcal{Q} := \{(\bx', \by') \in \mathbb{R}^{|T|} \times \mathcal{R}_{\by} : \| \by' - \bPhi_T \bx' \|_2 \text{ is minimized}\},
\end{equation*}
the set of minimizers for $\|\by' - \bPhi_T \bx'\|_2$. We also let
\begin{equation*}
(\tilde{\bx}, \tilde{\by}) = \argmin_{(\bx', \by') \in \mathcal{Q}} \| \by' - \by \|_2.
\end{equation*}
Then we define the functions:
\begin{equation*}
\text{resid}(\by, \bPhi_T) := \tilde{\by} - \bPhi_T \tilde{\bx},\quad \text{pcoeff}(\by, \bPhi_T) := \tilde{\bx},
\end{equation*}
which give the residual and projected coefficients, respectively.

These functions may be interpreted as projection operations onto the quantization region of $\by$, i.e. $\mathcal{R}_{\by}$. Note that $\tilde{\bx}$ and $\tilde{\by}$ may be uniquely determined by first solving the quadratic optimization problem,
\begin{equation*}
\tilde{\by} = \argmin_{\by' \in \mathcal{R}_{\by}} \| \by' - \by\|_2,
\end{equation*}
which gives a unique solution for $\tilde{\by}$, then solving for $\tilde{\bx}$ by
\begin{equation*}
\tilde{\bx} = \argmin_{\bx} \| \tilde{\by} - \bPhi_T \bx\|_2,
\end{equation*}
which also admits a unique solution.

Note that this problem is computationally tractable since the constraint $\by \in \mathcal{R}_{\by}$ is a set of linear inequalities. Though no theory has been developed for this algorithm, it has been shown to work well empirically. The algorithm is summarized below.  Here and throughout, we write $\supp_K(\bx)$ to denote the set of indices corresponding to the largest $K$ entries of $\bx$ in magnitude, $\supp_K(\bx) = \max_{|T|\leq K}\|\bx_T - \bx\|_2$.

\begin{algorithm}[H]
\caption{Quantized Subspace Pursuit (QSP)}\label{qsp}
\begin{algorithmic}
\STATE \textbf{Input:} sparsity level $K$, measurement matrix $\bPhi \in \R^{M \times N}$, compressed quantized signal $\by \in \R^M$
\STATE \textbf{Initialize:} $T^0 = \supp_K( \bPhi^* \by )$, $\by_r^0 = \text{resid}(\by, \bPhi_{T^0})$
\REPEAT
\STATE $\tilde{T}^l = T^{l - 1} \cup \supp_K( \bPhi^* \by_r^{l - 1}).$
\STATE $\bx_p = \text{pcoeff}(\by, \bPhi_{\tilde{T}^l})$ and $T^l = \supp_K(\bx_p)$
\STATE $\by_r^l = \text{resid}(\by, \bPhi_{T^l})$
\UNTIL{$\| \by_r^l \|_2 > \| \by_r^{l - 1} \|_2$}
\STATE $T^l = T^{l - 1}$
\STATE \textbf{Output:} $\hat{\bx}/\|\hat{\bx}\|_2$ where $\hat{\bx}_{\{1, ..., N\} - T^l} = 0$ and $\bx_{T^l} = \bPhi^{\dagger}_{T^l} \by$
\end{algorithmic}
\end{algorithm}

\subsection{Quantized Iterative Hard Thresholding (QIHT)}

The Quantized Iterative Hard Thresholding (QIHT) \cite{jacques2013quantized} algorithm is based on the Iterative Hard Thresholding (IHT) \cite{blumensath2009iterative} and Binary Iterative Hard Thresholding (BIHT) \cite{biht} algorithms. IHT was introduced for iteratively reconstructing a sparse signal in classical compressed sensing. The algorithm may be interpreted as solving:

\begin{equation*}
\begin{aligned}
\min_{\bx} \frac{1}{2}\| \by - \bPhi \bx \|_2^2 \quad\text{s.t.}\quad\| \bx \|_0 \leq K.
\end{aligned}
\end{equation*}

IHT solves this problem by iteratively computing
\begin{equation*}
\bx^{l + 1} = \eta_K \left( \bx^l + \bPhi^* \left(\by - \bPhi \bx^l\right)\right),
\end{equation*}
where $\eta_K(\bx)$ thresholds $\bx$ by maintaining the $K$ largest entries in magnitude of $\bx$ ($\supp_K(\bx)$) and setting the rest to zero. $\bx$ is initialized at $\bx^0 = 0$. This algorithm was shown to converge for $\|\bPhi\|_2 < 1$ in \cite{blumensath2009iterative}.

We may interpret the IHT algorithm as taking a gradient step to minimize the consistency-enforcing objective $\frac{1}{2}\| \by - \bPhi \bx \|_2^2$ then giving the best $K$-term approximation by hard thresholding.

Similarly, in the $1$-bit setting, BIHT modifies the gradient step in IHT and iteratively computes the following:
\begin{equation*}
\bx^{l + 1} = \eta_K \left( \bx^l + \mu \bPhi^* \left(\by - \sign(\bPhi \bx^l)\right)\right).
\end{equation*}
where $\mu$ is a scalar that controls the gradient step-size.

This may be interpreted as attempting to minimize the following objective:
\begin{equation}\label{biht}
\begin{aligned}
\min_{\bx\in S^{N-1}} \mu \| [\by \odot (\bPhi \bx)]_- \|_1 \quad\text{s.t.}\quad \| \bx \|_0 \leq K,
\end{aligned}
\end{equation}
where $\mu \in \mathbb{R}$ is the chosen gradient step size, $\odot$ is component-wise multiplication, and $[ ~ \cdot ~]_-$ is the projection of each component to the negative real line, i.e. for each component of $\bx \in \mathbb{R}^N$,
\begin{equation}\label{eqdef}
[x_i]_- =  \begin{cases}
x_i & \mbox{ if } x_i < 0\\
0 & \mbox{ otherwise.}
\end{cases}
\end{equation}

Note that the first term of the objective function enforces the inequality
\begin{equation*}
\by \odot (\bPhi \bx) \geq \boldsymbol{0},
\end{equation*}
which ensures the consistency of signs of $\by$ and $\bPhi \bx$.

Motivated by this optimization problem, \cite{jacques2013quantized} formulated a similar problem for multiple quantization values and thresholds by considering the following objective:
\begin{equation}\label{qihtobj}
\min_{\bx}  \sum_{k = 1}^M \sum_{j = 2}^{2^B} w_j |[\sign((\bPhi \bx)_k - \tau_j)(y_k - \tau_j))]_-| \quad \text{s.t.}
\quad \| \bx \|_0 \leq K,
\end{equation}
where $w_j = m_j - m_{j - 1}$. This optimization objective may be interpreted as the sum of the BIHT objective over all possible quantization thresholds. Calculating the subgradient of this function gives the update
\begin{equation*}
\ba^{l + 1} = \bx^l + \mu \bPhi^*\left(\by - f_Q(\bPhi \bx^l)\right).
\end{equation*}
which we then threshold by
\begin{equation*}
\bx^{l + 1} = \eta_K (\ba^{l + 1}).
\end{equation*}

QIHT is summarized below in Algorithm \ref{qiht} \cite{jacques2013quantized}.

\begin{algorithm}[H]
\caption{Quantized Iterative Hard Thresholding (QIHT)}\label{qiht}
\begin{algorithmic}
\STATE \textbf{Input:} measurement matrix $\bPhi \in \R^{M \times N}$, compressed quantized signal $\by \in \R^M$, quantization function $f_Q$, $\mu > 0$ step size, stopping criterion
\STATE \textbf{Initialize:} $\bx^0 = \frac{\bPhi^* \by}{\|\bPhi^* \by\|}$
\WHILE{not converged}
\STATE $\ba^{l + 1} = \bx^l + \mu \bPhi^*(\by - f_Q(\bPhi \bx^l))$
\STATE $\bx^{l + 1} = \eta_K ( \ba^{l + 1})$
\ENDWHILE
\STATE \textbf{Output:} $\hat{\bx} = \bx^k/\|\bx^k\|_2$
\end{algorithmic}
\end{algorithm}

Intuitively, the algorithm takes a subgradient step with step-size $\mu$ then projects the signal back to the $K-\ell_0$ sphere. Though the stability and convergence of QIHT have not yet been proven, numerical results and a limited case analysis suggest accurate empirical performance.

Note that when QIHT uses 1-bit measurements, it reduces to BIHT. When QIHT uses extremely fine measurements, the IHT method is recovered.

Like any method, QIHT is robust to noise that does not change the quantized values.  
To address noise that changes the quantization values of $\bPhi \bx$, we modify QIHT using Adaptive Outlier Pursuit, which we describe in the next section.

\section{New Adaptations}\label{novel greedy}

We present two new algorithm variants for reconstruction from quantized measurements.

\subsection{Quantized CoSaMP (QCoSaMP)}

Motivated by the adaptation of QSP, we consider the adaptation of a similar method, CoSaMP \cite{NeedeT_CoSaMP} to the quantized setting.   
Quantized Compressive Sampling Matching Pursuit (QCoSaMP) is described below.

\begin{algorithm}[H]
\caption{Quantized Compressive Sampling Matching Pursuit (QCoSaMP)}\label{qcosamp}
\begin{algorithmic}
\STATE \textbf{Input:} measurement matrix $\bPhi \in \R^{M \times N}$, quantized compressed signal $\by = \bPhi \bx$, sparsity level $K > 0$, maximum number of iterations $I$
\STATE \textbf{Initialize:} $\ba^0 = 0$, $\bv = \by$, $k = 0$
\WHILE{$k < I$}
\STATE Set $\bu = \bPhi \bv$, $\Omega =\supp_{2K}(\bu)$
\STATE Merge: $T = \Omega ~\cup~ \supp( \ba^{k-1})$
\STATE Projection: $\bx = \argmin_{\bu \in \mathcal{R}_{\by}} \| \by - \bu\|_2$\\ $\bb_T  = \argmin_{\bx}\|\bu - \bPhi_{T} \bx\|_2$ and $\bb_{T^c} = 0$
\STATE Set $\ba^k = \bb$
\STATE Update: $\bv = \by - f_Q(\bPhi\ba^k)$
\STATE $k = k+1$.
\ENDWHILE
\STATE \textbf{Output:} $\hat{\bx} = \ba^k/\|\ba^k\|_2$
\end{algorithmic}
\end{algorithm}

The primary difference between QCoSaMP and CoSaMP is the projection step; instead of estimating the signal using least squares, $\bPhi_T^{\dagger}\by$, we project  $\by$ into the quantization region $\mathcal{R}_{\by}$, and then apply least squares, as described in the algorithm above. 
Also, when computing the sample update, we include the quantizer $f_Q$ in computing the residual. 

We will see later that empirically QCoSaMP accurately reconstructs a signal from its quantized measurements.  Like the other approaches, theoretical guarantees may also be possible; we leave this for future work.

\subsection{Quantized Adaptive Outlier Pursuit (AOP-QIHT)}

In many applications, unintended noise may be added to the measurements during acquisition and transmission, creating both pre-quantization and post-quantization noise. Since QIHT's performance is best demonstrated when measurements are quantized into the correct corresponding bin, this motivates the use of Adaptive Outlier Pursuit following \cite{RefWorks:19} to detect potential quantization errors.

Mathematically, we can model our pre-quantization noise as $\bn$ where $(\bPhi \bx)_i + n_i$ for some fixed number of $i$. If $n_i$ is large enough in magnitude for some $i$, then it may cause $(\bPhi \bx)_i$ to be quantized in a different bin. Post-quantization noise may be modeled as sparse noise in which a measurement is quantized to a completely random quantization value.
In either case, we can model both forms of noise as some sparse vector $\bn$ such that its non-zero components satisfy
\begin{equation*}
f_Q\left((\bPhi \bx)_i + n_i\right) \neq f_Q\left((\bPhi \bx)_i\right).
\end{equation*}

Motivated by ideas from 1-bit compressive sensing using Adaptive Outlier Pursuit \cite{RefWorks:19} and QIHT \cite{jacques2013quantized}, we define
\begin{equation*}
\phi(x, y) = \sum_{j = 2}^{2^Q} \left|[(x - \tau_j)(y - \tau_j)]_-\right|.
\end{equation*}
 Given $x$ and $y$, $\phi(x, y)$ penalizes the case where $x$ and $y$ do not lie in the same quantization region.

We may then formulate our problem as
\begin{equation}\label{aop-qiht}
\begin{aligned}
 \min_{\bx\in S^{N-1}, \bn} \sum_{k = 1}^M  \phi\left((\bPhi \bx + \bn)_k, y_k \right)\quad \text{s.t.}\quad \| \bn \|_0 \leq L, \;\;\| \bx \|_0 \leq K. 
\end{aligned}
\end{equation}
where we penalize over all components of $\bPhi \bx + \bn$ and $\by$.

To solve this problem, we introduce a new binary variable, $\Lambda \in \{0, 1\}^M$, such that
\begin{equation*}
\Lambda_k = \begin{cases}
1, & \mbox{ if } y_k = f_Q((\bPhi \bx + \bn)_k) \\
0, & \mbox{ otherwise.}
\end{cases}
\end{equation*}

and redefine our problem as the following
\begin{equation*}
\min_{\bx\in S^{N-1}, \bLambda\in\{0,1\}^M} \sum_{k = 1}^M \Lambda_k  \phi((\bPhi \bx)_k, y_k)\quad \text{s.t.}\quad
 \sum_{k = 1}^M (1 - \Lambda_k) \leq L, \; \; \| \bx \|_0 \leq K. 
\end{equation*}

We note a few observations: Firstly, our sparsity constraint, $\| \bn\|_0 \leq L$, is equivalent to the constraint $\sum_{k = 1}^M (1 - \Lambda_k) \leq L$ by definition of $\Lambda$. 

Secondly, the objective for AOP-QIHT also differs slightly from QIHT's objective. Instead of weighing each inconsistent term by the same weight $w_j$, we allow the distance from each threshold to influence the weight of each term. Larger differences from each quantization threshold correspond to higher values of $\phi(y_i, (\bPhi \bx)_i)$, allowing for better detection of outliers.  

Lastly, if $L$ is unknown, which is true in many real-world applications, we must apply heuristics to choose $L$. If $L$ is chosen too small or too large, AOP-QIHT will perform worse. If $L$ is smaller, then corrupted measurements will stay in the measurements. If $L$ is larger, uncorrupted measurements will also classified as outliers leading to the loss of information.

Following M. Yan, et. al.'s solution for Adaptive Outlier Pursuit for BIHT \cite{RefWorks:19}, we apply the alternating minimization method:

\begin{enumerate}
\item Fix $\bLambda$ and solve for $\bx$:
\begin{equation}
\min_{\bx\in S^{N-1}} \sum_{k = 1}^M \Lambda_k \phi((\bPhi \bx)_k, y_k) \quad \text{s.t.}
\quad \| \bx \|_0 \leq K. 
\end{equation}

We can solve this by performing a QIHT update:
\begin{equation}
\bx^{l + 1} = \eta_K(\bx^l + \mu \bPhi^* (\by - f_Q(\bPhi \bx^l))).
\end{equation}

\item Fix $\bx$ and solve for $\bLambda$:
\begin{equation}\label{aop step}
\min_{\bLambda\in\{0,1\}^M} \sum_{k = 1}^M \Lambda_k \phi((\bPhi \bx)_k, y_k) \quad \text{s.t.}\quad \sum_{k = 1}^M (1 - \Lambda_k) \leq L.
\end{equation}

We can solve this by:
\begin{equation}
\Lambda_k = \begin{cases}
0 & \mbox{ if } \phi((\bPhi \bx)_k, y_k) \geq $M$\\
1 & \mbox{ otherwise}
\end{cases}
\end{equation}
where $M$ is the $L$th largest component of $\{\phi((\bPhi \bx)_k, y_k)\}_{k = 1}^M$.\\

\end{enumerate}

Intuitively, AOP adaptively removes outliers from the data that it assumes has been corrupted by noise by removing the $L$ largest terms from the objective. These terms are then not used in the subsequent update of the algorithm. Our proposed method is described as follows.

\begin{algorithm}[H]
\caption{Adaptive Outlier Pursuit for Quantized Iterative Hard Thresholding (AOP-QIHT)}\label{aop}
\begin{algorithmic}
\STATE \textbf{Input:} measurement matrix $\bPhi \in \R^{M \times N}$, quantized compressed signal $\by$, sparsity level $K > 0$, number of wrongly detected measurements $L>0$, $\mu > 0$ step size, maximum number of iterations $I > 0$ 
\STATE \textbf{Initialize:} $\bx^0 = \frac{\bPhi^* \by}{\|\bPhi^* \by\|}$, $k = 0$, $\Lambda = \boldsymbol{1} \in \bR^M$, $T = \{1,\ldots,M\}$, tol $=\infty$ , TOL $=\infty$ 
\WHILE{$l\leq I$ and $L \leq$ tol}
\STATE Compute $\ba^{l+1} = \bx^k + \mu \bPhi^*_T(\by - f_Q(\bPhi_T \bx^k))$
\STATE Update $\bx^{l+1} = \eta_K(\ba^{l+1})$
\STATE Set tol = $\|\by - f_Q(\bPhi \bx^{l+1})\|_0$
\IF{tol $\leq$ TOL}
\STATE Compute $\bLambda$ as above
\STATE Update: $T = \supp(\Lambda)$
\STATE Set TOL = tol.
\ENDIF
\STATE $l = l+1$.
\ENDWHILE
\STATE \textbf{Output:} $\hat{\bx} = \bx^k/\|\bx^k\|_2$
\end{algorithmic}
\end{algorithm}

\section{Numerical Experiments}\label{numerical experiments}

In this section, we perform several numerical experiments to demonstrate the effectiveness of the approaches. Furthermore, we perform an in-depth comparison and analysis of all the methods to choose a preferred bit-depth and algorithm for given noise, bit budget, and sparsity.

To setup these experiments, we generate a measurement matrix $\bPhi \in \R^{M \times N}$ whose elements follow an i.i.d. Gaussian distribution, a $K$-sparse signal $\bx^*$ whose non-zero entries are drawn from a standard Gaussian distribution which is then normalized to have unit norm. We compute our compressed signal $\by = f_Q(\bPhi \bx^*)$.

\subsection{QCoSaMP Experiments}

\begin{figure}[H]
\begin{center}
\includegraphics[scale=.5]{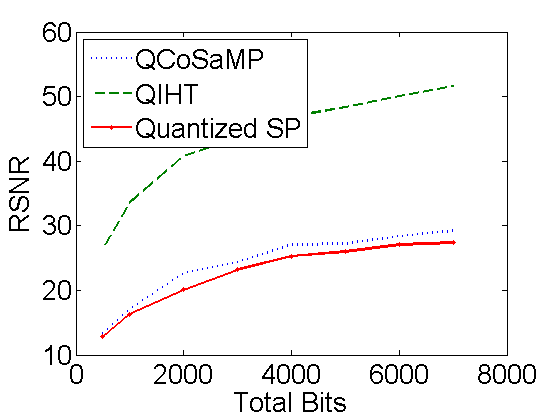}
\includegraphics[scale=.5]{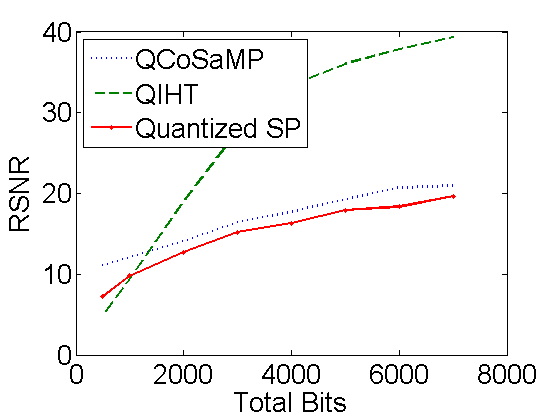}
\end{center}
\caption{Comparison of QCoSaMP, Quantized SP, and QIHT. We graph total bits against RSNR with $N = M = 1000$ and $K = 10$, averaged over 40 trials. The left uses 1-bit measurements and the right uses 4-bit measurements.}\label{fig1}
\end{figure}

In the first experiment, we set the signal size to be $N = 1000$ with sparsity $K = 10$, set $M=1000$ and vary the total bit budget $T_B$ and the bit-depth $B$ (bits per measurement, $B\approx \log(Q)$).  We compute the reconstruction SNR\footnote{We use $\text{SNR} = 10 \log \left( \frac{\| \bx^*\|_2^2}{\| \bx - \bx^*\|_2^2} \right)$.} (RSNR) over $40$ trials and compare the average for each method. 
The results are shown in Figure \ref{fig1}.  This experiment demonstrates that QIHT outperforms the other greedy approaches for extremely low bit-depth.  However, at higher bit-depths, QCoSaMP yields better recovery in the low bit budget regime.  A theoretical understanding of this phenomena is interesting future work. 




\subsection{AOP-QIHT Experiments}

\begin{figure}
\begin{center}
\includegraphics[scale=.5]{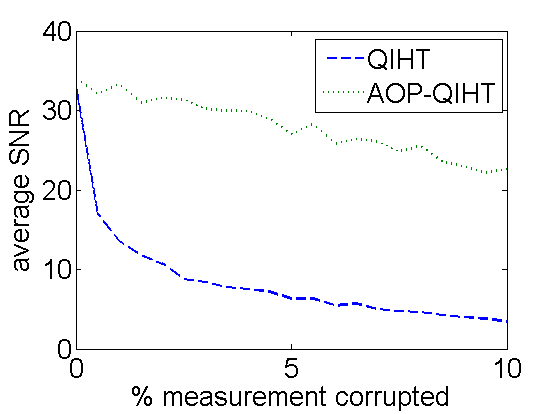}
\includegraphics[scale=.5]{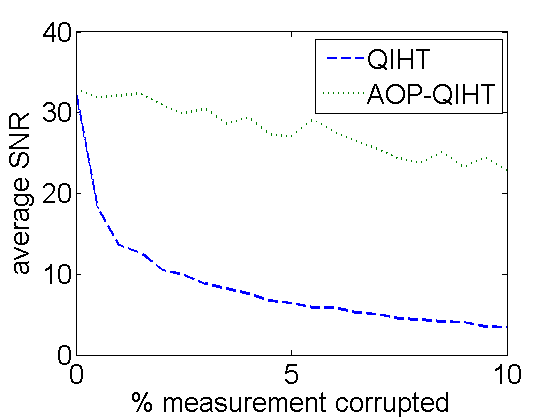}
\end{center}
\caption{Comparison of QIHT and AOP-QIHT on corrupted data with different noise levels. In these experiments, $N = M = 1000$ and $K = 10$. The top uses 1-bit measurements and the bottom uses 4-bit measurements. We plot the percentage of corrupted measurements against the average SNR over 40 trials.}\label{aopfig}
\end{figure}

In this experiment we keep the parameters as above, but in each trial we select a few measurements and flip their sign to corrupt the measurement's quantization.  We vary the corruption level to be between 0\% and 10\% of the total number of measurements and record the average SNR over 40 trials. 
Figure \ref{aopfig} demonstrates that AOP-QIHT outperforms QIHT for all noise levels, especially when noise affects a higher percentage of measurements.  We observe that AOP-QIHT performs significantly better than QIHT especially for $>2\%$ measurement corruption. For future work, we can also examine how this algorithm performs for different quantization schemes and different types of noise.




\subsection{Quantized CS Algorithm Comparison and Analysis}

In this set of experiments, we compare all the greedy algorithms and examine which algorithms and bit-depths perform best for a given sparsity, total bits, and noise level. We consider sparsity $K = \{2, 4, 6, 8, 10, 12, 14, 16 \}$, total bits $T_B = \{500, 1000, 2000, 3000, 4000, 5000, 6000, 7000, 8000, 9000, 10000\}$, input signal-to-noise ratio $ISNR = \{35, 20, 10\}$, and bit depth $B = \{1, 2, 3, 4\}$. We average over 20 trials.  We show results for selected parameter values, with omitted results that vary as one expects in between those we show here.  In addition, we summarize all results compactly in Figures \ref{figcool} -- \ref{figcool3}.  These figures catalog which algorithm exhibits the best performance for various parameter choices.  We hope that this serves as a useful tool for practitioners when deciding which method to use for their application.

We observe that for lower noise levels, QIHT and AOP-QIHT tend to perform better while QCoSaMP and Quantized SP perform better for higher noise. This may be explained by the hard thresholding that QIHT and AOP-QIHT performs. AOP-QIHT also performs better than QIHT for higher noise. Most graphs also indicate that 1-bit is best.




\begin{figure}
\begin{center}
\includegraphics[scale=.5]{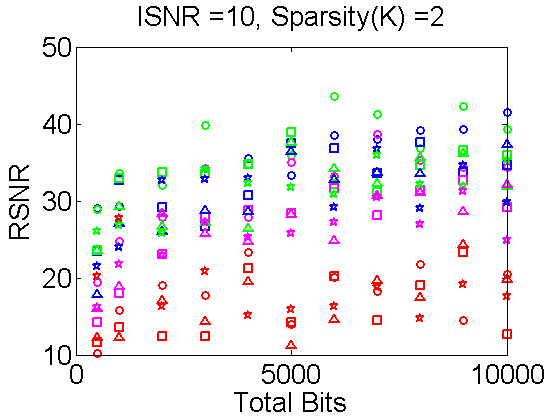}
\includegraphics[scale=.5]{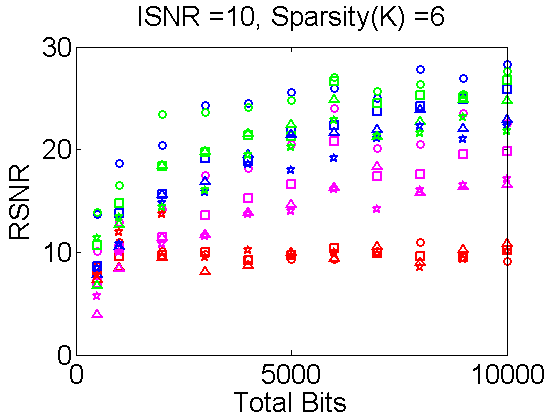}
\includegraphics[scale=.5]{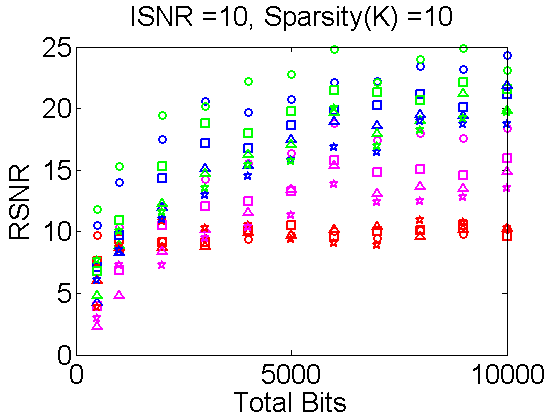}
\includegraphics[scale=.5]{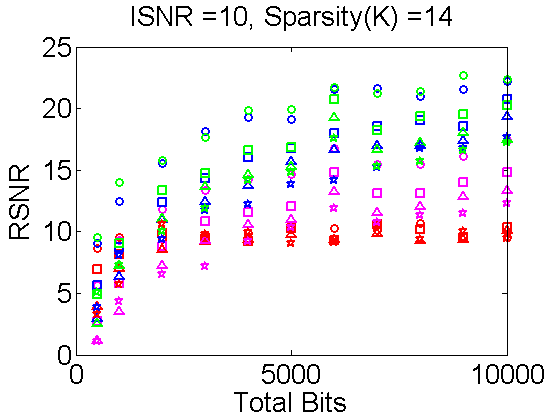}
\end{center}
\caption{Comparison of QIHT (red), AOP-QIHT (magenta), QCoSaMP (green), and Quantized SP (blue) with bit-depths 1 (circle), 2 (square), 3 (triangle), and 4 (star). These figures graph total bits against RSNR for ISNR = 10 and various sparsity levels. }
\end{figure}








\begin{figure}
\begin{center}
\includegraphics[scale=.5]{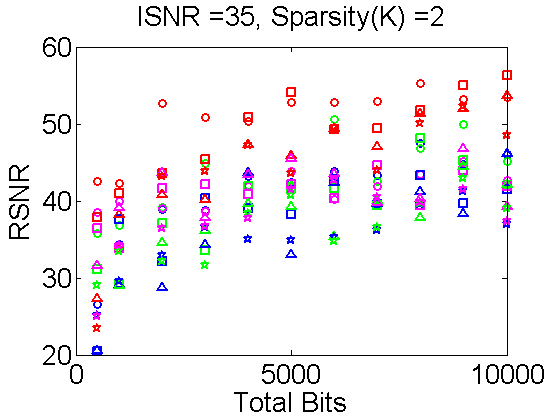}
\includegraphics[scale=.5]{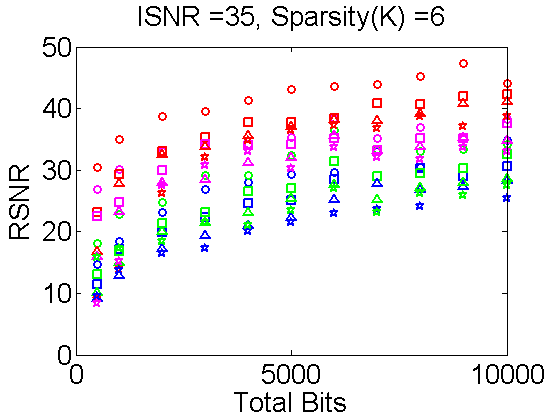}
\includegraphics[scale=.5]{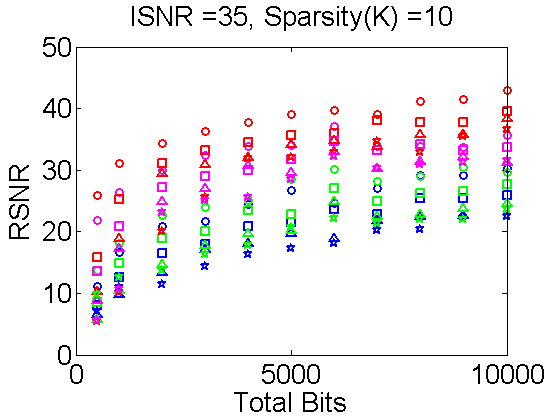}
\includegraphics[scale=.5]{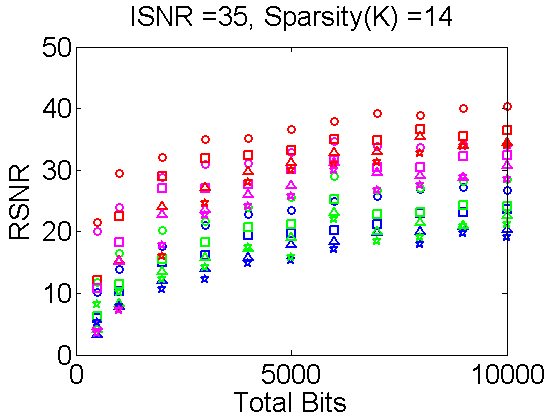}
\end{center}
\caption{Comparison of QIHT (red), AOP-QIHT (magenta), QCoSaMP (green), and Quantized SP (blue) with bit-depths 1 (circle), 2 (square), 3 (triangle), and 4 (star). These figures graph total bits against RSNR for ISNR = 35 and various sparsity levels. }
\end{figure}




\begin{figure}
\begin{center}
\includegraphics[scale=.5]{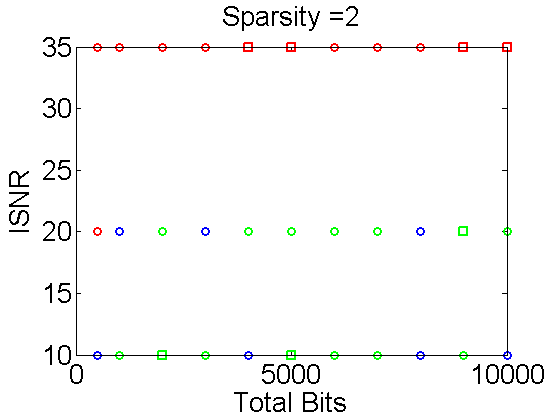}
\includegraphics[scale=.5]{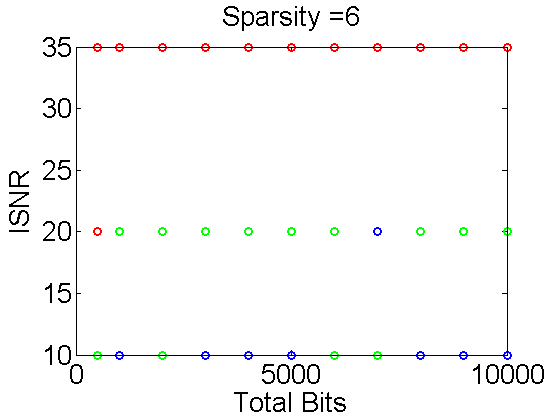}
\includegraphics[scale=.5]{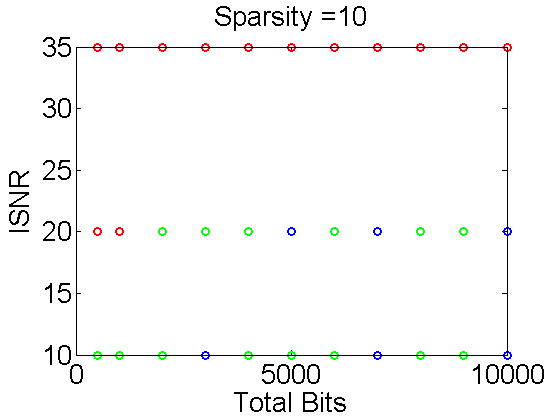}
\includegraphics[scale=.5]{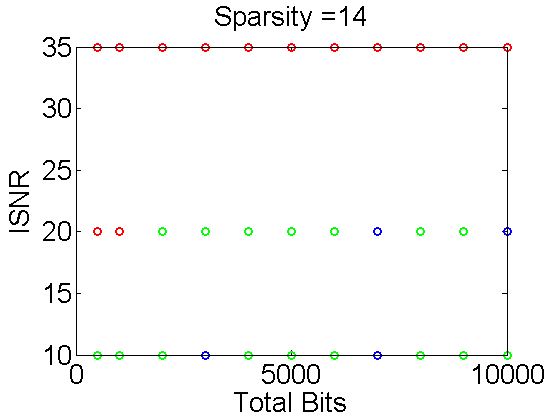}
\end{center}
\caption{Comparison of QIHT (red), AOP-QIHT (magenta), QCoSaMP (green), and Quantized SP (blue) with bit-depths 1 (circle), 2 (square), 3 (triangle), and 4 (star). These figures graph the best algorithm and bit-depth for given total bits and ISNR for various sparsities.}\label{figcool}
\end{figure}




\begin{figure}
\begin{center}
\includegraphics[scale=.5]{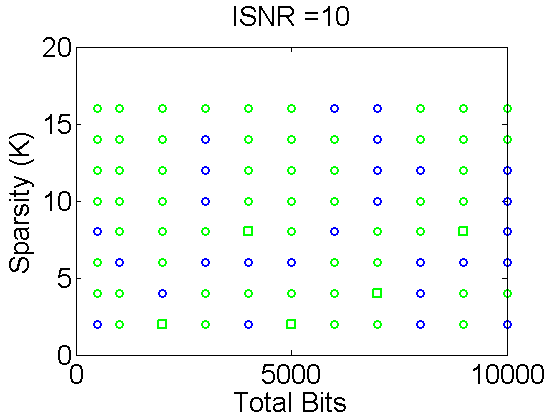}
\includegraphics[scale=.5]{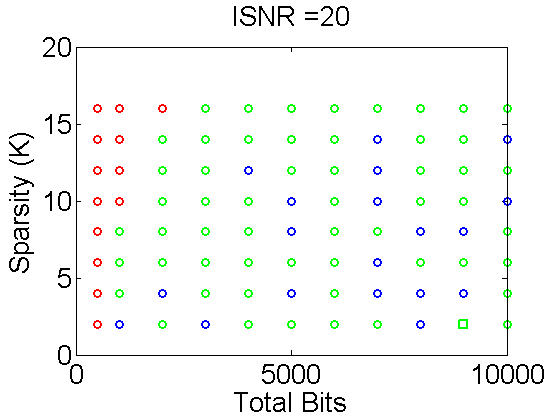}
\includegraphics[scale=.5]{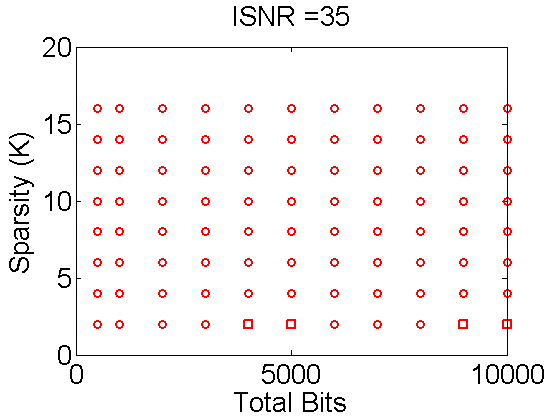}
\end{center}
\caption{Comparison of QIHT (red), AOP-QIHT (magenta), QCoSaMP (green), and Quantized SP (blue) with bit-depths 1 (circle), 2 (square), 3 (triangle), and 4 (star). These figures graph the best algorithm and bit-depth for given total bits and sparsity for various ISNR.}\label{figcool2}
\end{figure}




\begin{figure}
\begin{center}
\includegraphics[scale=.5]{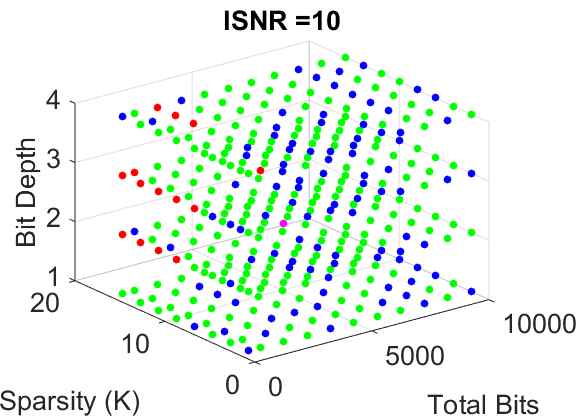}
\includegraphics[scale=.5]{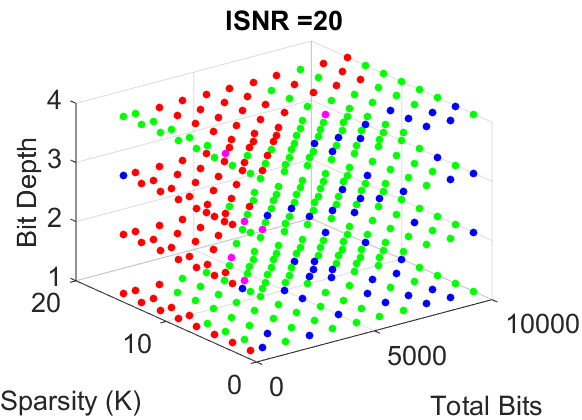}
\includegraphics[scale=.5]{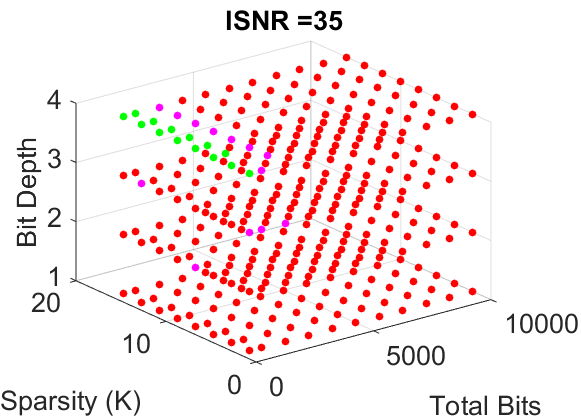}
\end{center}
\caption{Comparison of QIHT (red), AOP-QIHT (magenta), QCoSaMP (green), and Quantized SP (blue). These figures graph the best algorithm for given total bits, sparsity, and bit-depth for various ISNR.}\label{figcool3}
\end{figure}

\section{Conclusion} \label{conclusion}
Compressed sensing has furnished a rigorous theory and class of methods for signal reconstruction from compressed measurements.  However, for practical use one must also consider how quantization affects the reconstruction error.  Moreover, extreme quantization -- where each measurement is captured by only a few bits -- can be useful in its own right due to the efficiency of implemented hardware. 

In this paper, we propose two novel robust greedy algorithms for reconstructing quantized compressed signals. QCoSaMP modifies CoSaMP by projecting onto the quantization region by solving a set of computationally tractable optimization problems, then considering quantization in the residual. AOP-QIHT iteratively detects falsely quantized measurements and recovers signals from the ``correct" measurements by applying QIHT. 

Allowing the bit-depth and bit-budget to be additional parameters in the CS framework, one now needs to select a method from the existing approaches that is most in line with the desired parameter regime.  For that reason, in this paper we compare the performance of four greedy algorithms, Quantized SP, QCoSaMP, QIHT, and AOP-QIHT over multiple settings of parameters for normalized signals.  We show that the one-bit QIHT and AOP-QIHT algorithms tend to perform best for low-noise cases while QCoSaMP and QSP perform better for higher noise.  We believe this is a useful tool that will help guide practitioners when navigating the existing approaches.

\section*{Acknowledgements}

The work of Hao-Jun Michael Shi and Mindy Case was supported in part by the California Research Training Program for Computational and Applied Mathematics 2015 under NSF Grant DMS $\#1045536$. The work of Xiaoyi Gu and Shenyinying Tu was supported in part by the California Research Training Program for Computational and Applied Mathematics 2015 under {NSF CAREER $\#1348721$}. The work of Deanna Needell was supported under {NSF CAREER $\#1348721$}.  The authors would also like to thank Prof. Andrea Bertozzi for hosting the summer program in which this work originated.

\bibliographystyle{myalpha}
\nocite{*}
\bibliography{bibl}

\end{document}